\documentclass[prb,10pt,  twocolumn,superscriptaddress,amssymb]{revtex4-1}
\usepackage{graphicx}% Include figure files
\usepackage{dcolumn}% Align table columns on decimal point
\usepackage{bm}% bold math
\usepackage{tabularx}
\usepackage{bm}%
\usepackage{color}      % use if color is used in text
\usepackage[colorlinks=true,linkcolor=blue,citecolor=blue,urlcolor=blue]{hyperref}%

\begin{document}

\title{Efficiently Engineered Room Temperature Single Photons in Silicon Carbide}
\author{S. Castelletto}
\email{stefania.castelletto@mq.edu.au}
\affiliation{%
Diamond Science, Department of Physics and Astronomy, Macquarie University, Sydney, NSW 2109, Australia
}%

\author{B. C. Johnson}
\email{johnson.brett@jaea.go.jp}
\affiliation{%
Radiation Effects Group, Japan Atomic Energy Agency, 1233 Watanuki, Takasaki, Gunma 370-1292, Japan
}%
\author{N. Stavrias}
\affiliation{%
School of Physics, The University of Melbourne,Victoria 3010, Australia
}%
\author{T. Umeda}
\affiliation{%
Graduate School of Library, Information and Media Studies, University of Tsukuba, Tsukuba 305-8550, Japan
}%

\author{T. Ohshima}
\affiliation{%
Radiation Effects Group, Japan Atomic Energy Agency, 1233 Watanuki, Takasaki, Gunma 370-1292, Japan
}%

\begin{abstract}
\begin{center}
\line(1,0){250}
\end{center}

{\bf We report the first observation of stable single photon sources in silicon carbide (SiC). These sources are extremely bright and operate at room temperature demonstrating that SiC is a viable material in which to realize various quantum information, computation and photonic applications. The maximum single photon count rate detected is 700k counts/s with an inferred quantum efficiency around 70\%. The single photon sources are due to intrinsic deep level defects constituted of carbon antisite-vacancy pairs. These are shown to be formed controllably by electron irradiation. The variability of the temporal kinetics of these single defects is investigated in detail.}
 \begin{figure} [h!]
 \includegraphics[width=14.2cm]{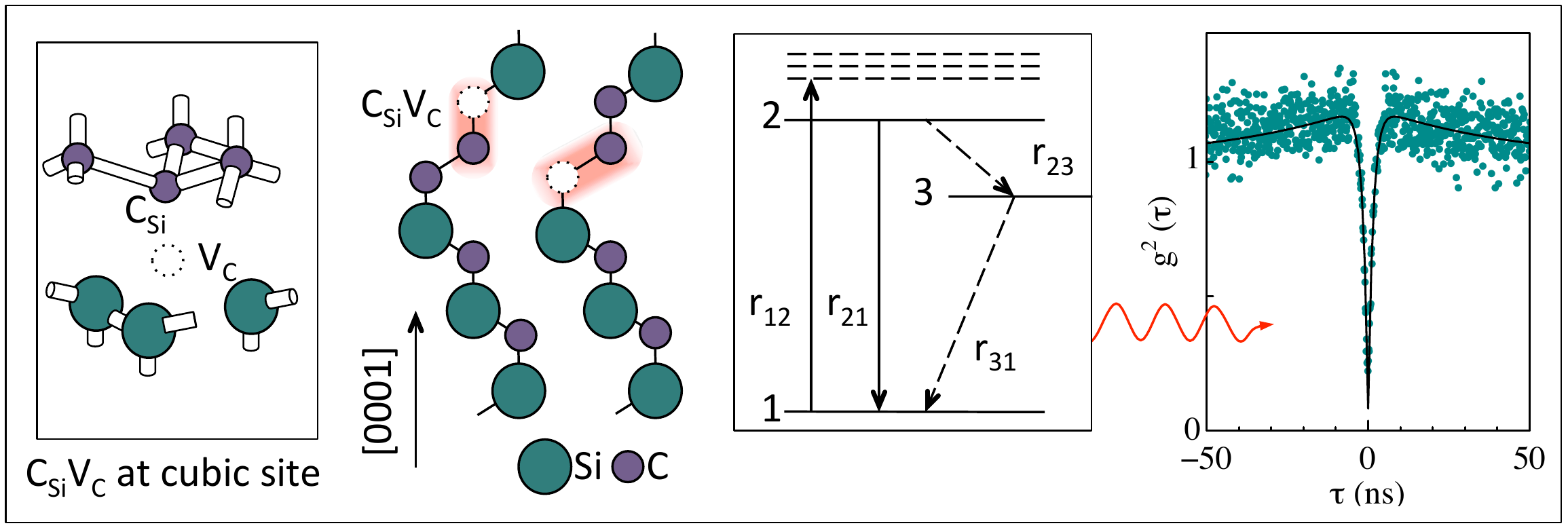}
 \end{figure}

\end{abstract}

\maketitle

Silicon carbide has fast become an important wide-bandgap semiconductor for high-power, high-temperature applications and is regarded as a key material for next generation photonic\cite{Song11,Yamada2011} and electronic devices\cite{Madar04,Sei-Hyung98}. As such, SiC has well-established growth and device engineering protocols\cite{Nakamur04}. This has allowed the development of a mature platform for nano-fabrication and high-Q photonic crystal cavities, which have already been reported for a spectral region from 550 to 1450~nm\cite{Yamada2011}. SiC is considered one of the best biocompatible materials, especially to blood \cite{Melinon2007}, and the synthesis of SiC nano-particles have been employed as fluorescent biological labels in cell imaging\cite{Fan08}.

Such versatility has led to recent interest in the possibility of using SiC as a platform for various photonic, spintronic and quantum information applications\cite{OBrien2009}. In particular, it is a promising material in which to host qubits and thus holds great promise for the future implementation of scalable quantum processors\cite{Weber2010,DiVincenzo}. In particular, several authors\cite{Weber2010,Koehl2011,Baranov2011} indicate that SiC could harbor equivalent defects to the nitrogen vacancy center in diamond, a well known solid-state qubit\cite{Balasubramanian}. While candidate ensemble optically-active defects, having the net angular momentum (or spin) required to act as qubits operating at room temperature, have been identified\cite{Son2006,Koehl2011,Baranov2011}, single photon sources and the ability to engineer them controllably have not yet been demonstrated in SiC.

In this work, we present, to the best of our knowledge, the first observation and engineering of single photon sources from point defects in SiC. At room temperature the defect responsible emits between 670--710~nm and is assigned to the carbon antisite-vacancy pair\cite{Steeds2009,Umeda2006}. A great number of these single defects can be created, indicating that the electron irradiation results in a very efficient method to engineer single defects in SiC. The study of the kinetics and photoluminescence (PL) of these single defects using confocal microscopy and single photon correlation, reveals a variability of the photophysics of the defects, that is undetectable with ensemble measurements. As the identification of deep defects in wide bandgap semiconductors is typically carried out by photoluminescence and electron paramagnetic resonance ensemble measurements\cite{Umeda2006}, the direct correlation to single defects is often difficult. Here, we perform a study of the defect formation from the ensemble level down to the single defect level, providing a reliable route for the identification of the defect responsible for the observed single photon source. We also correlate the carbon antisite-vacancy pair to the formation of another interesting defect for quantum information, the silicon vacancy in SiC\cite{Kolesov2012}.

The discovery of single photon emission in this material and its efficient creation, is the starting point to pioneer a new branch of alternative defects for next generation solid state candidates in quantum information, computation and photonics.

%%%%%%%%%%%%%%%%%
\section*{Results}

\subsection*{Single defect measurements}

\begin{figure*}[t!]  %[h!]
\includegraphics[width=12.5cm]{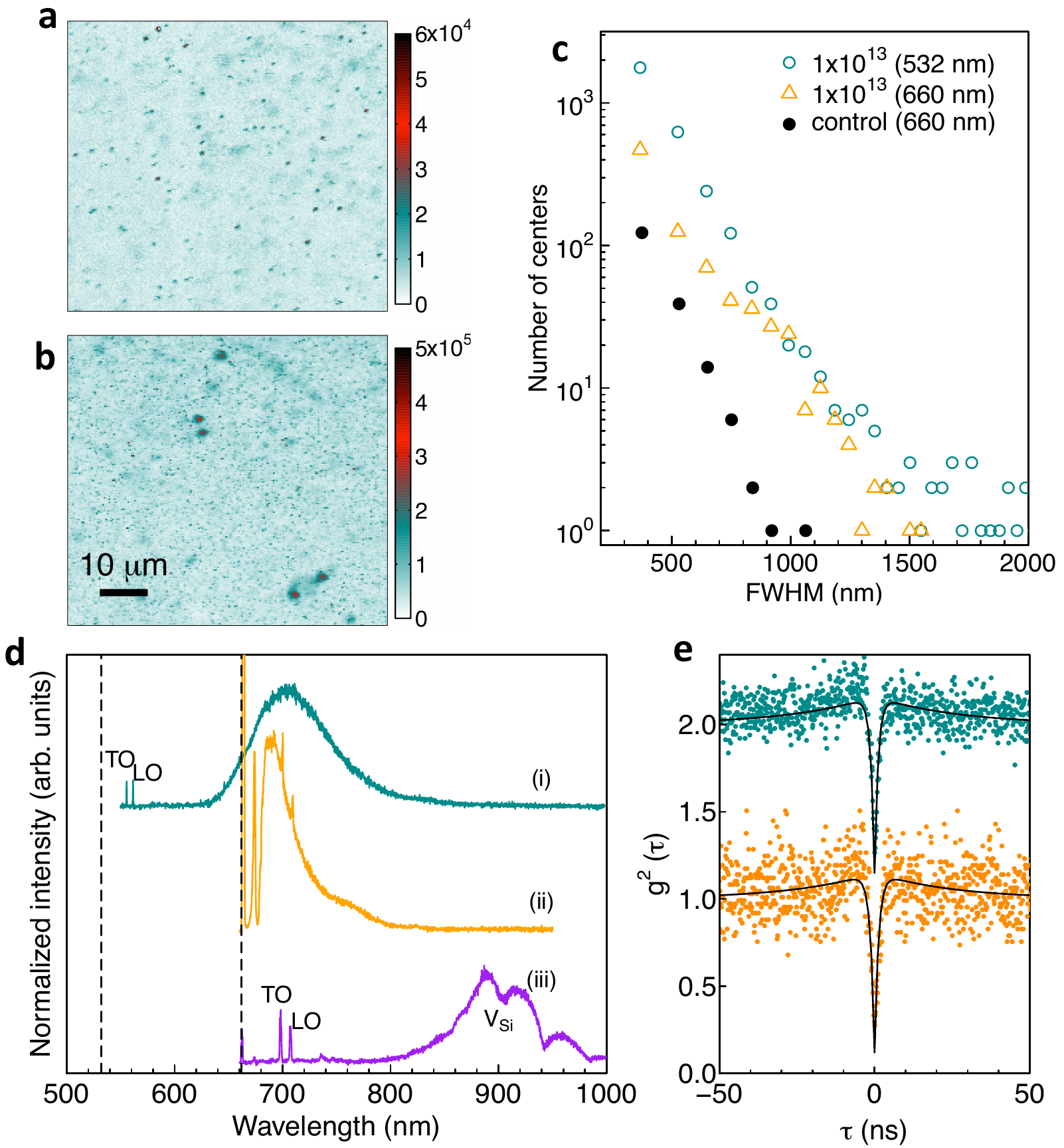}
\caption[]{{\bf Confocal microscopy images and spectra of single photon sources in SiC.} Confocal map of the low electron fluence sample excited with 660~nm ({\bf a}) and 532~nm ({\bf b}) lasers. The diffraction limited spots all correspond to single photon emitters. Map (a) has been scanned using a 10~nm FWHM 694~nm filter. Note the difference in color scale. {\bf c,} Histogram of the number of centers observed as a function of the spatial FWHM measured and counted from ({\bf a,b}). The histogram from a confocal map of an unprocessed sample is also shown indicating a clear increase in the defect density after irradiation and annealing at 300$^{\circ}$C. {\bf d,} Normalized room temperature PL of a single photon source in the irradiated sample excited with 532~nm (i) and 660~nm (ii,iii) lasers. The laser wavelengths are indicated by the dashed lines. {\bf d}(iii), PL of a center which was not anti-bunching. The transverse and longitudinal optic Raman modes (TO, LO) are also indicated. {\bf e,} The corresponding anti-bunching curves for {\bf d}(i) and (ii).}
\label{f1}
\end{figure*}

Figure~\ref{f1}a,b shows confocal PL maps of a high purity, semi-insulating SiC wafer irradiated with 2~MeV electrons and subsequently annealed (see Methods: Samples preparation). Spatially isolated, extremely bright defect centers are observed, the number and brightness of which are dependent on the excitation wavelength as shown in the histogram of Fig.~\ref{f1}c. It is clear that the 532~nm excitation laser allowed a greater variety of defects to be addressed. As observed in the confocal maps in Fig.~\ref{f1}, higher count rates are detected when using the 532~nm excitation laser, mainly due to the use of an ultra-steep long pass filter with 90\% transmittance so that the single photons were collected on their entire spectral bandwidth. Centers could also be observed in untreated sample but at a concentration of almost an order of magnitude lower than the processed sample (Fig. \ref{f1}c). The concentration of these centers was found to increase monotonically with electron irradiation fluence and increasing anneal temperature (see Supplementary). Using a Hanbury-Brown and Twiss interferometer to measure the photon correlation we found that all the diffraction limited spots with diameters of around 300--400~nm demonstrated single photon emission with wavelengths between 670--710~nm.

Typical PL spectra of these bright centers are shown in Fig.~\ref{f1}d excited using both 532 and 660~nm lasers. The width of the PL emission tended to be narrower when excited with a 660~nm laser. However, the PL line-shape possesses some variability both at 532 nm and 660 nm excitation as shown in the Supplementary Information. Remarkably, photon correlation spectroscopy revealed that centers emitting single photons were the dominant defect, i.e. the second order photon correlation function at zero delay is mostly $g^{(2)}(0) < 0.5$ (Fig.~\ref{f1}e).

\begin{figure*}[t!]  %[h!]
\includegraphics[width=12.5cm]{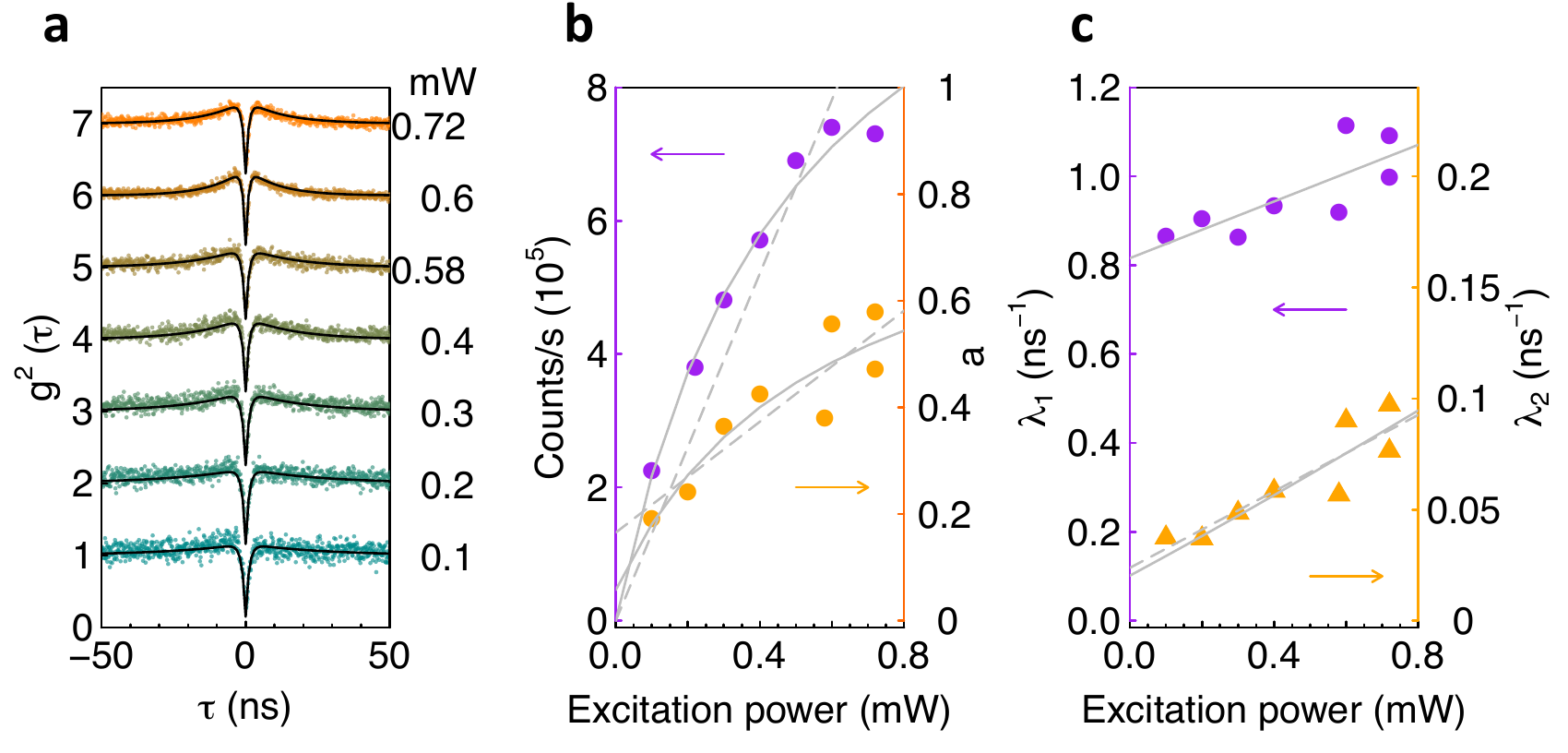}
\caption[]{{\bf Temporal dynamics of single photon sources in SiC.} {\bf a,} Second-order correlation function at different 532~nm excitation powers from the center with PL shown in Fig.~\ref{f1}d(i). A weak dependence of the metastable state shelving with the optical excitation power is observed. The curves are shifted vertically for clarity. {\bf b,} Saturation curve of the defect fitted with Eq.~\ref{phi} where $\beta = 2.2\;\rm mW^{-1}$ (solid line) and $\beta = 0$ (dashed line), to infer the quantum yield (left axis). $a$ versus excitation power, fitted with the $\beta$ term (solid) line and with $\beta = 0$ (dotted line) (right axis). {\bf c,} Decay rates, $\lambda_1$ and $\lambda_2$ extracted from the fits in {\bf a} versus the excitation power fit with the model with the $\beta$ term (solid line) and without it (dashed line).}
\label{f2}
\end{figure*}

The photo-physics and the population dynamics from the ground state to the excited state of these single photon sources was investigated. The PL temporal dynamics can be studied by analyzing the $g^{(2)}(\tau)$ at different excitation optical powers. Fig.~\ref{f2}a shows the anti-bunching behavior of the center with PL shown in Fig~\ref{f1}d(i) for a range of 532~nm excitation powers. Due to the presence of a photon-bunching effect (higher correlation at longer delay times), each curve was fit with a three-level system model,

\begin{equation}
\label{g2}
g^{(2)}(\tau)=1-(1+a) \exp[- \lambda_1 \tau]+ a \exp[- \lambda_2 \tau]
\end{equation}

\noindent where the decay rates, $\lambda_1$ and $\lambda_2$ are free parameters and related to the transition rates, $r_{ij}$, from level ($i$) to level ($j$) by the equations $\lambda_1= r_{12}+r_{21}$, $\lambda_2= r_{31}+r_{23} r_{12}/ \lambda_1$ \cite{Kitson}. The values of $a$, is a measure of the photon-bunching, that prevents radiative decay to the ground state (1) from the excited state (2) by a metastable state (3) and can be expressed as $a= r_{12} r_{23}/( \lambda_1 r_{31})$. Photon-bunching at longer delay times in single molecules\cite{Basche92} and color centers\cite{Kurtsiefer00} in diamond has been associated with the presence of such a long-lived metastable state. Non-radiative decay to the ground-state takes place from this metastable state, inducing an overall reduction of the single photon emission arising from low rate electron trapping in the metastable state.

For this particular center the saturation count rate and optical saturation power are $\Phi_{\infty}=1.3 \times 10^{6} \rm\; counts/s$ and $P_{sat} = 0.5$~mW, respectively. This is obtained from fitting the data in Fig.~\ref{f2}b with $\Phi=\Phi_{\infty} P_{opt}/( P_{sat} + P_{opt})$. The dependence of the transition rate parameters of this model, $a$ and $\lambda_{1,2}$, on the excitation power are shown in Fig.~\ref{f2}(b,c). A linear optical power dependence of the coefficient $\lambda_1= r_{21}^{0}(1+ \alpha P_{opt})$ is assumed, where $r_{12}= \alpha r_{21}^{0}P_{opt}=\sigma I_{opt} /(h \nu)$, is related to the absorption cross-section of the defect, $\sigma$. Due to the very weak dependence of the metastable state population on the excitation power, we have assumed that the shelving ($r_{23}$) and de-shelving ($r_{31}$) decay rates of the metastable state are constant. Other models used to describe the population dynamics of the excited state and of the metastable state versus the excitation optical power\cite{Aharonovich2010,Neu12}, however, may also describe our data. For example, in some cases a linear dependence of $r_{31}=r_{31}^{0}(1+\beta P_{opt})$ may be employed where $\beta$ is a fitting parameter that describes the variation of the decay rate from the metastable to ground states with changes in optical excitation power. This parameter is important since it signifies that a power dependent effect in de-shelving the metastable level is operative. This dependence has also been observed for single molecules\cite{Treussart01}. Fits with and without the $\beta$ term are shown in Fig.~\ref{f2}(b,c).

On fitting the three-level system model, we obtain the decay rate terms $r_{21}^{0}$, $r_{31}^{0}$, $r_{23}$, $\alpha$ and $\beta$.  Extrapolation to zero excitation power gives the excited state lifetime, $r_{21}^{0}$, and the shelving decay rates, $r_{31}$. While extrapolation to infinite excitation power of $\lambda_{2}$ gives $r_{23}$. For the center shown in Fig.~\ref{f2} we obtain $1/r_{21}^{0}=(1.2 \pm 0.1)$~ns, $\alpha = (0.3\pm0.1)$, $\beta= (2.1\pm0.5)$~mW$^{-1}$, $1/r_{31}^{0}=(45 \pm 3)$~ns, $1/r_{23}^{0}=(7 \pm 2)$~ns, and for $\beta= 0$, $1/r_{31}^{0}=(23 \pm 2)$~ns, $1/r_{23}^{0}=(8 \pm 2)$~ns.  The absorption cross section is $\sigma \approx 5 \times 10^{-16} \;\rm cm^{2}$. Using these values the following model can be used to determine the quantum efficiency, $\eta_{qe}$, of this single photon source,

\begin{equation}
\label{phi}
\Phi = \eta_{qe} \eta_{coll} \eta_{det} r_{21}(1+ r_{21}/ r_{12}+ r_{23}/ r_{31})^{-1}.
\end{equation}

\noindent By fitting the count rate we obtain $\eta_{qe} \eta_{coll} \eta_{det} =0.008\pm0.001$. If we assume that the detection efficiency is only limited by the detector's quantum efficiency ($\eta_{det}$=0.55) with the filters having a transmission of 90\%, and a collection efficiency of our setup determined to be $\eta_{coll}\approx$ 0.02\%\cite{Castelletto11}, $\eta_{qe}\approx$ 70\% is estimated.

To assess the properties of the single photon emission accurately, the variability of their kinetics must be established. Therefore, many photo-stable emitters were measured with the two different excitation wavelengths and analyzed.  We observed a moderate dispersion of the excited state lifetime from 1.2--2.1~ns, yielding average values of $1.5\pm0.3$~ns (20\% variability) and $\alpha = (0.4\pm0.3)\;\rm mW^{-1}$. Concerning the metastable state, due to the very small photon-bunching effect observed with increased excitation power, the measurement of the decay rates from and to the metastable state are affected by a large variability, giving average values of $1/r_{31}^{0}=(40\pm20)$~ns and $1/r_{23}=(13\pm5)$~ns. In addition not all the emitters were best fit with the $\beta$ parameter.

The saturation properties of the defects also display a remarkable variability, with optical saturation powers observed between 0.2 to 3.3~mW and saturation count rates found from $2.5\times 10^{5}$~counts/s to the maximum observed of $2\times 10^{6}$~counts/s.

The observed SiC defects have a much shorter metastable state lifetime compared to typical diamond color centers (of the order of 100 ns for NV or $\mu s$ for other defects such as Cr-related or SiV centers)\cite{Kurtsiefer00,Neu12,Aharonovich2010}. In addition, the here observed large dispersion of its lifetime could indicate that the photon-bunching effect in the correlation function may be also attributed to a longer-lived non-radiative state with different charge or to thermal excitation from the excited states to higher energy non-radiative states. These states can either belong to the center or they can be formed by other impurities in the local nanoenvironment, as it occurs in some diamond defects. Low temperature and longer delay time photon correlation measurements are required to fully identify the photon-bunching associated physics.

\subsection*{Single photon source stability}

\begin{figure*}[t!]  %[h!]
\includegraphics[width=12.5cm]{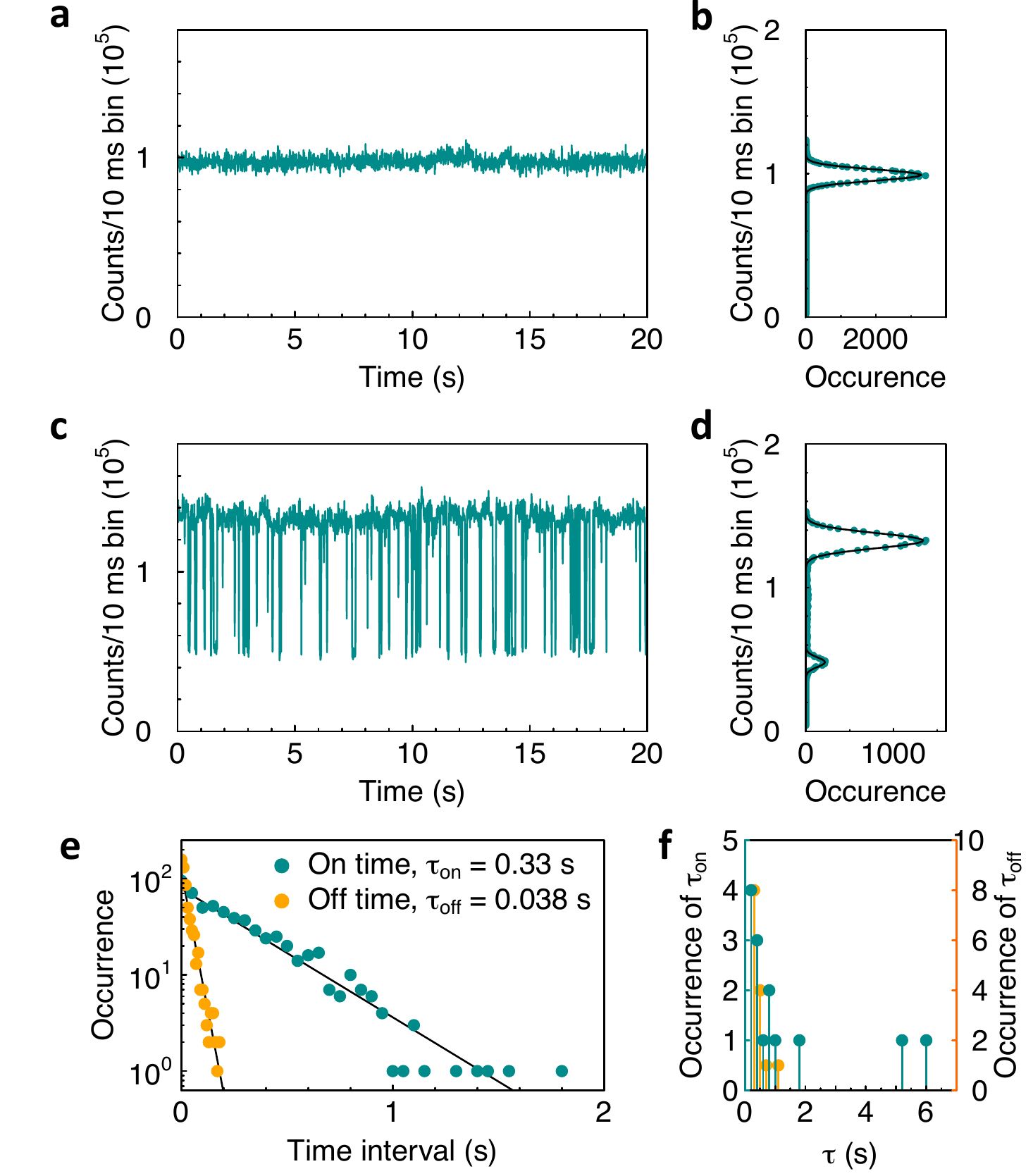}
\caption[]{{\bf Photo-stability of single photon sources in SiC.} {\bf a}, 20~s fragment of a PL time-trace sampled into 10~ms bins from a stable single photon source with its corresponding  count rate histogram ({\bf b}). {\bf c,d}, A similar time trace is shown for a blinking center. The on and off times of ({\bf b}) on a semi-logarithmic scale. {\bf f,} Histogram of the measured $\tau_{on}$ and $\tau_{off}$ of the 55 emitters analyzed. 45\% of the emitters in this sample were stable and did not show blinking behavior.}
\label{f3}
\end{figure*}

We studied the PL emission stability of these defects and observed that 45\% of the emitters measured in the irradiated and 300$^{\circ}$C annealed sample are stable over a long period of time. An example of a center with a stable PL time trace is shown in Fig.~\ref{f3}a along with its associated Gaussian count rate histogram Fig.~\ref{f3}(b). Interestingly, the remaining centers show PL blinking with some eventually photo-bleaching, even if not permanently. Such centers in samples annealed at higher temperatures (500$^{\circ}$C) were almost absent suggesting that the blinking was related to the density of non-radiative impurities which could be annihilated with annealing. An emitter with on-off fluorescence statistics, between two PL states is shown in Fig.~\ref{f3}(c). By measuring all on/off-time durations in a single trace, the corresponding probability of occurrence $P(\tau_{on,off})$ can be calculated\cite{Bradac2010}. The distribution of the ``on'' and ``off'' times are plotted on a semi-logarithmic scale in Fig.\ref{f3}(e) together with the fits. Both $P(\tau_{on})$ and $P(\tau_{off})$ show an exponential trend following the equation:

\begin{equation}
P(\tau_{on,off} ) \propto \exp[- \tau/(\tau_{on,off})]
\end{equation}

\noindent with the time constant, $\tau_{on}$ usually much greater than $\tau_{off}$, especially in unprocessed SiC or samples annealed at higher temperatures. This model suggests that typically a single mechanism is operative, which could involve a photo-induced charge conversion of the addressed center due to defects in close proximity acting as charge traps or donors. $P(\tau_{on,off})$ is also found to have a strong laser power dependence supporting a photo-induced mechanism (see Supplementary). Two state intermittence is the most common blinking observed. However, a small number of centers displayed more complex PL blinking. This was especially true when using 532~nm laser excitation where three states or emission with no stable level could be observed from a single anti-bunching center. Centers excited with the 660~nm laser showed greater photo-stability.

\subsection*{Ensemble measurements}
\begin{figure*}[t!]  %[h!]
\includegraphics[width=12.5cm]{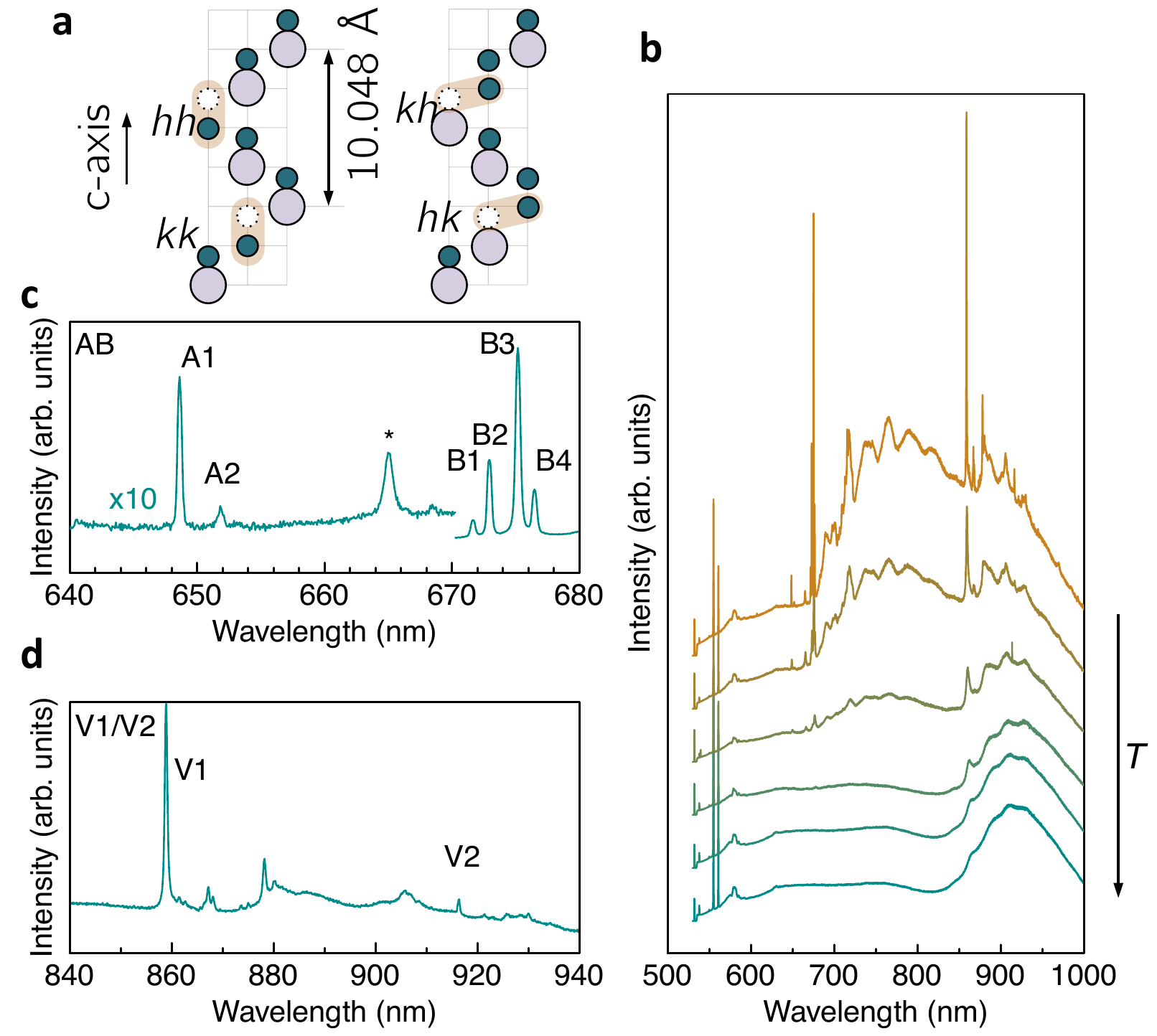}
\caption[]{{\bf Ensemble measurements of defects in electron irradiated SiC.}  {\bf a,} Structure of the four possible $\rm C_{Si}V_{C}$ type defects in SiC shown in the (1120) crystallographic plane. Small and large circles represent C and Si atoms while the small open circles are vacancies. $h$ and $k$ denote the hexagonal and cubic sites in the lattice, respectively. {\bf b,} PL of a sample irradiated with 2 MeV electrons to a fluence of 1$\times$ 10$^{17}$ cm$^{-2}$ at temperature of 80~K (top spectrum) to room temperature (bottom spectrum). Expanded view of the 80 K PL spectra showing the $\rm C_{Si}V_{C}$ ({\bf c}) and V$_{\rm Si}$ ({\bf d}) defect signatures. The asterisk in ({\bf c}) indicates a signal of unknown origin.}
\label{f4}
\end{figure*}

Ensemble room temperature and low temperature measurements were performed on a sample irradiated to $10^{17}\;\rm e/cm^{2}$ at 532~nm using a Renishaw Micro-Raman spectrometer. At 80~K the so called AB lines can clearly be observed (Fig.~\ref{f4}b,c) and result in a broad PL band at room temperature. These lines are associated with the neutral (C$\rm _{Si}$V$\rm _{C}$)$^0$ defect \cite{Steeds2009}. The atomic model of this defect with its four possible cubic or hexagonal orientations is shown in Fig.~\ref{f4}a. The single photon emissions observed in this work are attributed to this particular defect. The 532~nm laser excites all zero phonon lines. The 660~nm laser excites only the B spectra and the phonon side band is detected.

Similarly, the V1 and V2 lines\cite{Baranov2011} are also observed, corresponding to defects on inequivalent sites of the V$\rm _{Si}$ defect (Fig.~\ref{f4}b,d). These two defect types, $\rm C_{Si}V_{C}$ and V$\rm _{Si}$, are the dominant defects observed in this study. Confocal mapping revealed that the V1 and V2 lines are dominant and have a higher intensity than the AB lines in the high electron fluence samples. In samples irradiated with lower fluences this defect was rarely observed due to the greater threshold energy required for the Si sub-lattice of SiC. We did not observe anti-bunching in the V$\rm _{Si}$ center possibly due to the dominance of the C$\rm _{Si}$V$\rm _{C}$ defect. Excitation with wavelengths closer to the V1/V2 system may be required to observe single photon emission even in un-processed samples.

%%%%%%%%%%%%%%%%%
\section*{Discussion}

These experiments have demonstrated the formation of stable single photon sources in SiC operating at room temperature. The emission around 700~nm can be modeled with a three-level system containing a metastable state with a short lifetime compared to single photon sources in diamond. The brightness of this single photon source is comparable or even higher than the brightest color centers observed in bulk diamond\cite{Aharonovich2010}. Whilst as grown defects yielding single photon emission are observed, scaling of the number of defects with irradiation fluence, indicate that the defects can be successfully engineered. Unusual blinking behaviour is observed which is possibly due to non-radiative recombination at other defects sites in close-proximity to the addressed center. Blinking could be reduced when a high annealing temperature was employed. The defect responsible for the single defect emission is identified as the carbon anti-site vacancy pairs in its neutral charge state, ($\rm C_{Si} V_{C}$)$^0$. The deployment of these single centers  and their further characterization is extremely promising in quantum information applications and may trigger further studies of point defects in SiC at the single defect level.

%%%%%%%%%%%%%%%%%%%%%%%%%%%%%%
%%%%%%%%%%%%%%%%%%%%%%%%%%%%%%
\section*{Methods}
\subsection*{Sample preparation}

The material under study is high purity semi-insulating (HPSI) SiC purchased from CREE. According to the manufacturer, the as-received samples contained B (10$^{14}$ cm$^{-3}$), C interstitials (10$^{14}$ cm$^{-3}$) and N ($<$10$^{14}$ cm$^{-3}$).

Samples were irradiated with 2 MeV electrons to fluences of 1$\times$10$^{13}$, 1$\times$10$^{15}$ and 1$\times$$^{17}$  e/cm$^2$  at a flux of 2.82$\times$10$^{12}$ cm$^{-2}$s$^{-1}$. The samples were held at temperatures below 80$^{\circ}$C during irradiation in a nitrogen ambient to inhibit Frenkel pair annihilation and reduce the possibility of contamination. After irradiation, a low temperature (300$^{\circ}$C) anneal in an Ar ambient for 30 minutes was performed. This temperature was chosen so as to enhance the concentration of various defects of interest based on ensemble electron spin resonance (ESR) measurements\cite{Umeda2006}. An unprocessed sample from the same wafer was also characterized in order to establish the effectiveness of the irradiation procedure employed.

Similar samples were also annealed for 30 minutes at 300, 400, 500$^{\circ}$C in an Ar ambient after 2~MeV electron irradiation to a fluence of $10^{13}\rm\; e/cm^{2}$.

\subsection*{Photoluminescence and photon correlation measurements.}
A Hanbury-Brown-Twiss interferometer was used to identify optically active single defects and measure the time correlation of the center emission. In the confocal microscope the samples were excited using 0.2-1.5 mW continuous wave laser operating at 532 nm (Coherent, model: Compass 315M-100) and continuous diode laser operating at 665 nm. The laser was focused onto the sample using a 100X infinity-corrected oil immersion objective lens with a NA of 1.4  and the luminescence was collected confocally through a 50~$\mu$m pinhole. A spectrometer (Princeton Instruments Acton 2500i, Acton) with a cooled CCD (Camera Pixis 100 model: 7515-0001 Princeton Instruments) was used to characterize the luminescence and a HBT interferometer with single-photon-sensitive avalanche photodiodes (Perkin Elmer SPCM-AQR
14) was used to measure the photon statistics. Photon counting and correlation was carried out using a time-correlated single-photon-counting (TCSPC) module (PicoHarp 300, PicoQuant GmbH).

The SiC samples were mounted on a piezo XYZ stage allowing 100x100~$\mu m^{2}$ scans. The unwanted residual laser line was eliminated by a 660 nm long pass filter. A filter with 694$\pm$10 nm was placed in front of HBT interferometer to observe single photons during 660 nm laser excitation. The spectroscopy was done using only a long pass filter. All the single photon/single defect measurements were performed at room temperature.

For ensemble PL measurements we used a Renishaw MicroRaman Spectrometer equipped with a Linkam stage for temperature control. The excitation source was a 532~nm laser with 60~mW of power at the back of the objective. The objective was a 50x long working distance with 0.5~NA. A 532~nm edge filter was used. The spectrometer has a working distance of 0.25m and uses an 8$\mu$m slit and a 1200~g/mm grating. The detector is a fan cooled Si CCD. We used 10~s accumulations per point.

\section*{Acknowledgements}
SC acknowledges the funding support from the Australian Research Council Centre of Excellence EQuS. This work is supported by the Japanese Society for the Promotion of Science (JSPS) (Grant-in-aid for Scientific Research, 22.00802). This study is also partially supported by the Ministry of Education, Science, Sports and Culture, Grant-in-Aid for Challenging Exploratory Research, 2012, 24656025.

\section*{Author contributions}
SC and BCJ proposed the idea and the strategy for the experimental design and data analysis and wrote the paper. SC performed the optical single photon characterization experiments. BCJ irradiated and annealed the samples. NS performed the low temperature PL measurements. SC and BCJ coordinated the experiments and analyzed the data. SC, BCJ, NS, TO and TU discussed the results, and commented on the manuscript.

\section*{Additional Information}
The authors declare no competing financial interests. Supplementary information
accompanies this paper at www.nature.com/naturephotonics. Reprints and permission
information is available online at http://npg.nature.com/reprintsandpermissions. Reprints
and permissions information is available at www.nature.com/reprints. Correspondence
and requests for materials should be addressed to S.C (stefania.castelletto@mq.edu.au).

\newpage

\begin{center}
\begin{figure*}[t!]
\normalsize
{\large\bf Supplementary: \\ Efficiently Engineered Room Temperature Single Photons in Silicon Carbide}\\
\vspace*{1cm}
S. Castelletto,$^{1}$ B. C. Johnson,$^{2}$ N. Stavrias,$^{3}$  T. Umeda,$^{4}$  and T. Ohshima$^{2}$ \\
\vspace*{4pt}
{\it  $^{1}$Diamond Science, Department of Physics and Astronomy, \\ Macquarie University, Sydney, NSW 2109, Australia \\
$^{2}$Radiation Effects Group, Japan Atomic Energy Agency, \\ 1233 Watanuki, Takasaki, Gunma 370-1292, Japan \\
$^{3}$School of Physics, The University of Melbourne,Victoria 3010, Australia \\
$^{4}$Graduate School of Library, Information and Media Studies, \\ University of Tsukuba, Tsukuba 305-8550, Japan}
\vspace*{4pt}
\begin{center}
\line(1,0){250}
\end{center}
\end{figure*}
\end{center}

\vspace*{24cm}

\renewcommand{\thefigure}{S\arabic{figure}}
\setcounter{figure}{0}

\section*{Photoluminescence variability}

\begin{figure}[b!]  %[h!]
\includegraphics[width=8cm]{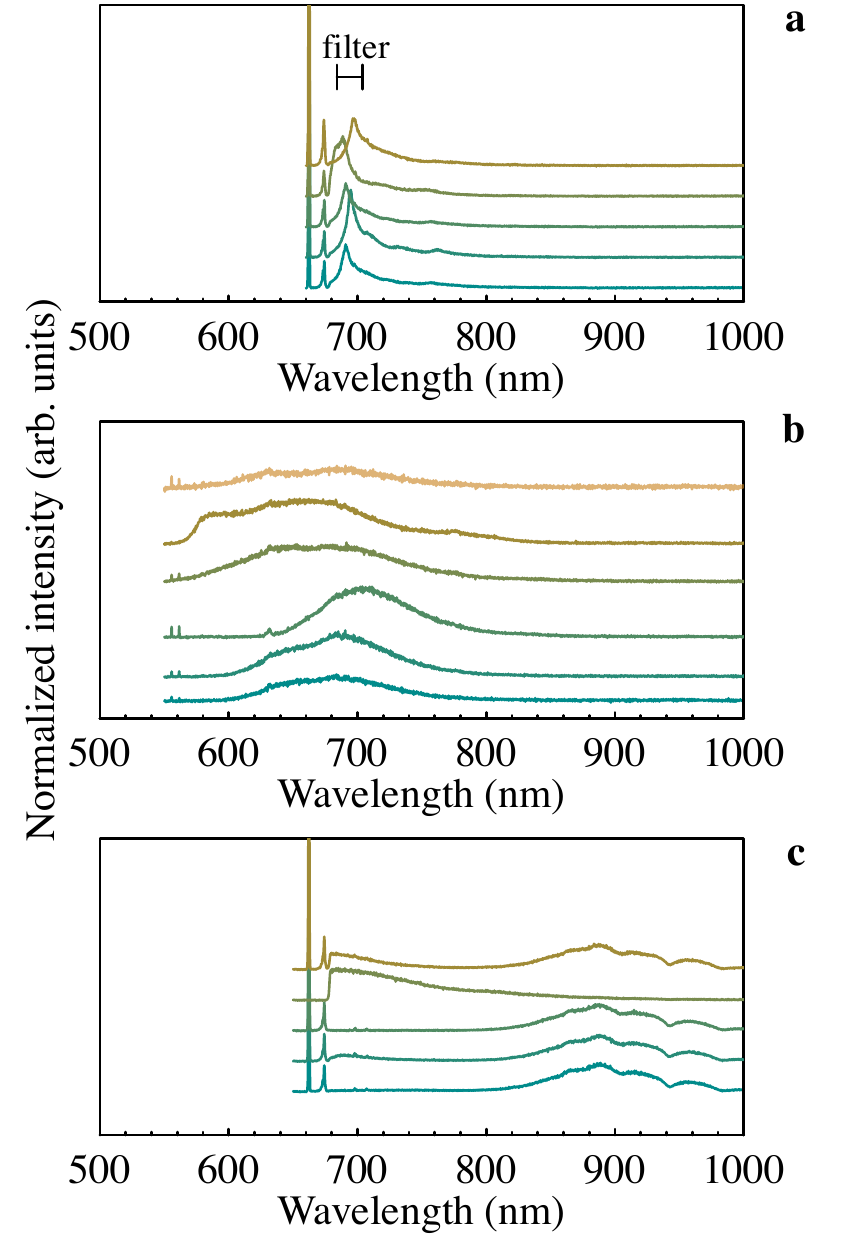}
\caption[]{{\bf Variability observed in PL emission at room temperature using confocal microscopy}. PL emission associated with the single photon center excited with 660~nm (a) or 532~nm (b) in samples irradiated with 10$^{13}$ e/cm$^2$. (c) PL emission measured in the sample irradiated to a fluence of 10$^{17}$ e/cm$^2$, excited at 660~nm. These emissions do not correspond to single defects.}
\label{sf1}
\end{figure}

We have used single defect spectroscopy to study the photo-physical parameters of single defects in SiC not accessible with conventional ensemble techniques such as standard electron paramagnetic resonance. A detailed analysis of the variability of these parameters is made possible. Firstly, the photoluminescence (PL) spectra of single photon sources in SiC is shown in Fig.~\ref{sf1}. The PL of samples irradiated with 2~MeV electrons to a fluence of 10$^{13}$ e/cm$^2$ and annealed at 300$^{\circ}$C are shown excited with either a 660~nm laser or a 532~nm laser are shown in Fig.~\ref{sf1}(a) and (b), respectively. Broad PL is observed between 600-800~nm. Fig.~\ref{sf1} (c) shows typical spectra from a sample irradiated to a fluence of 10$^{17}$ e/cm$^2$ followed by a similar anneal. The emission clearly corresponds with the V$_\mathrm{Si}$-related emission shown in Fig.~\ref{f4} of the main article. The emission related to the (C$_\textrm{Si}$V$_\textrm{C}$)$^0$ shows much greater temperature quenching but the PL band observed around 600-800~nm is associated with this defect and is responsible for the the single photon emission. There are in fact six different zero phonon lines associated with this defect in SiC which contribute to the variability of the band peak position.

\setcounter{page}{1}

\section*{Blinking statistics}
\begin{figure*}[t!]  %[h!]
\includegraphics[width=14cm]{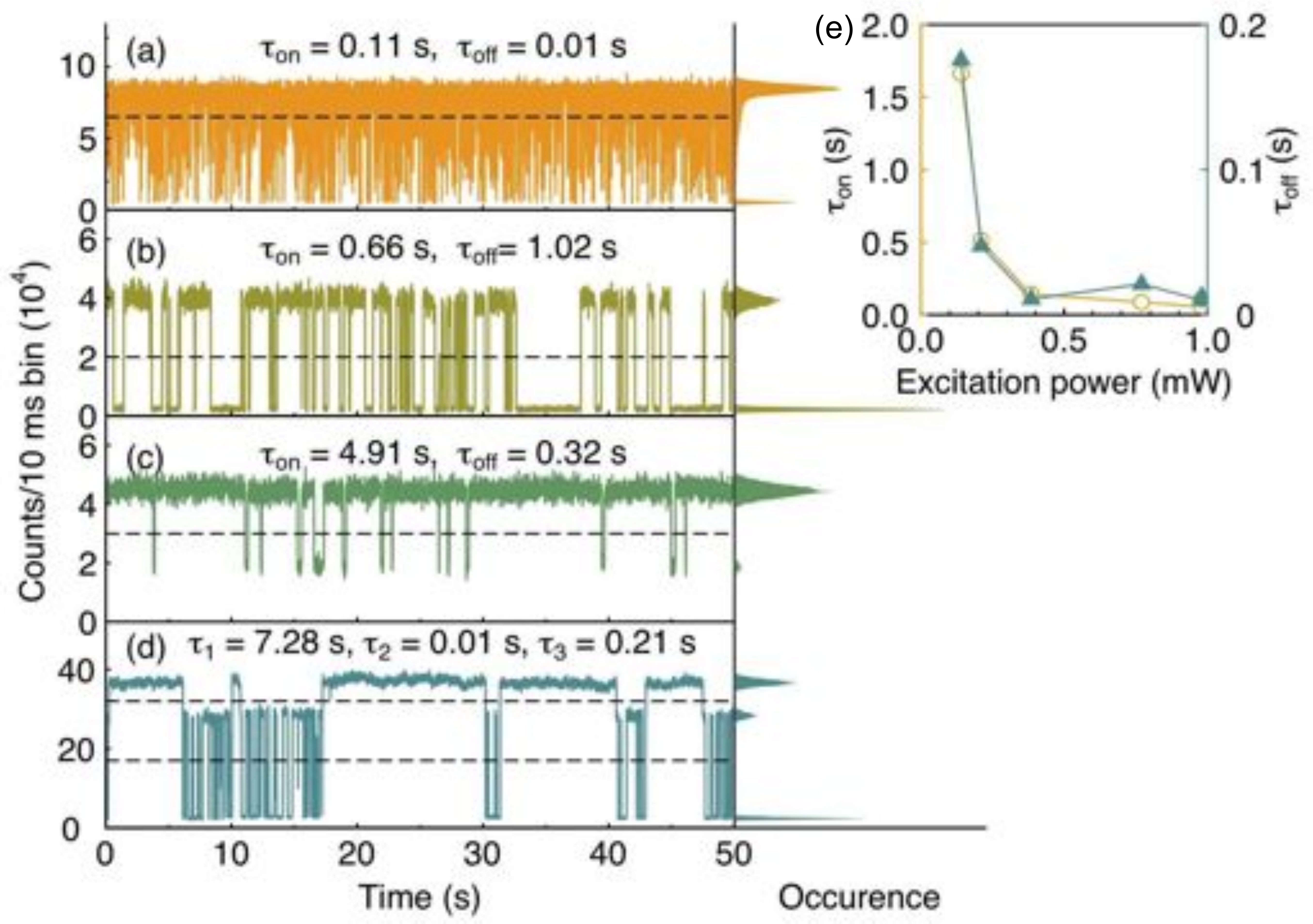}
\caption[]{{\bf Example of blinking statistics of single defects.} a-d show examples of blinking of single photon emitters in a SiC sample irradiated with 2~MeV electrons to a fluence of 10$^{13}$ e/cm$^2$ and annealed at 300$^{\circ}$C. The time constants for each state are also given. e, The on/off time constant dependence excitation power.}
\label{sf3}
\end{figure*}

We studied the statistical behavior of the ``on'' and ``off'' time distributions of single emitters and identified a number of blinking categories. One common case is an unclear blinking behavior as shown in Fig.~\ref{sf3}(a). Emitters having two on-off states are shown in Fig.~\ref{sf3}(b) and (c), where in case (b) the ``on'' time dominates the ``off'' time. The opposite is true in (c) where the ``off'' state has a non-zero intensity. Finally, Fig.~\ref{sf3}(d) shows blinking between 3 separate states. This last example is less common but is more likely to occur when 532 nm laser is used for excitation. Fig.~\ref{sf3}(e) shows the dependence of the ``on'' and ``off'' times on the excitation power for a two state single blinking centre, both the``on'' and ``off''  times reduce with increasing excitation power.

\section*{Defect concentration versus irradiation fluence and annealing temperature}

Using confocal imaging, we studied the concentration of the defects in samples prepared with different electron irradiation fluences $10^{13}$,  $10^{14}$ and  $10^{17}\;\rm e/cm^{2}$. For this purpose we used an excitation at 660 nm. We compared the results obtained with the PL observation to make sure we were observing the same defects. In addition, we also imaged a sample of the same SiC purity and transparency to determine preexisting as grown single defects. As expected the material contains the defects as grown with single photon emission signature, however we show that the number of defects scale with the electron irradiation fluence, indicating that the defect responsible for the single photon emission can be induced from the electron irradiation.

\begin{figure*}[b!]  %[h!]
\includegraphics[width=\linewidth]{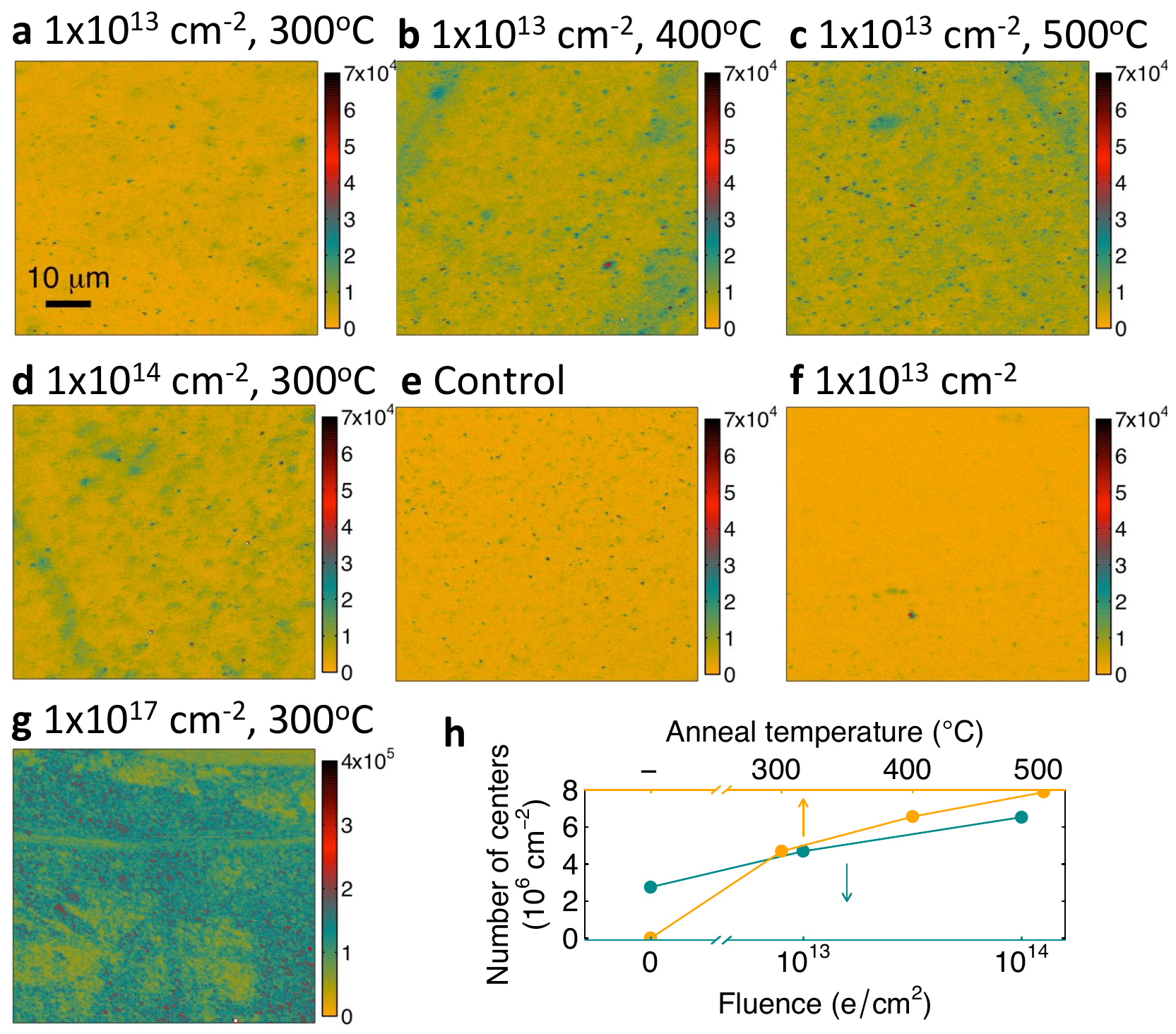}
\caption[]{{\bf Confocal maps with defects concentration versus irradiation fluence and annealing temperature.} {\bf a-g} Confocal maps of SiC with the processing conditions indicated. Note that only {\bf g} is on a different color scale since it is relatively bright and shows no isolated defect PL. {\bf h} The number of diffraction limited optical centers as a function of the anneal temperature (top axis) and fluence (bottom axis).}
\label{sf4}
\end{figure*}

We measured the concentration of the defects in samples prepared with the same electron irradiation fluence of $10^{13} \;\rm e/cm^{2}$ , applying different annealing temperatures (as described in Methods). For this purpose we again excited at 660 nm. We checked that the bright spots in the confocal were corresponding to single emitters with the same obtained PL emission. We observed a remarkable increase in the number of defects for higher temperature annealing. A summary of the total number of defects in a $100\times 100\;\rm  \mu m^{2}$ scan versus temperature and fluences is shown in the Fig.\ref{sf4}(h).

The increase in the number of defects with the increased annealing temperature is indicating that the defects studied here are more stable after high annealing temperature. In the 500 $^\circ$C annealing sample some defects were also appearing to be more than 1 ($g^{(2)}(0)\sim 0.5$), and very few defects were blinking, even if photo-bleaching was still observed.

\end{document}